\documentclass[reprint,amsmath,amssymb,aps,superscriptaddress,nofootinbib,prd]{revtex4-2}
\usepackage{graphicx}
\usepackage{dcolumn}
\usepackage{bm}
\usepackage[utf8]{inputenc}
\usepackage{amsmath, amsthm, amsfonts, amssymb}
\usepackage[svgnames]{xcolor}
\definecolor{DarkRed}{RGB}{179, 27, 27}
\colorlet{color1}{NavyBlue}
\usepackage[colorlinks=true,allcolors=DarkRed]{hyperref}
\usepackage{bm}
\usepackage{physics}
\usepackage{dsfont}
\usepackage{graphicx}
\usepackage{cleveref}
\usepackage{xfrac}
\usepackage{comment}
\usepackage{tikz}
\usepackage{pgfplots}
\usepgfplotslibrary{fillbetween}

\def\prd{{Physical Review D}}
\def\prl{{Phys. Rev. Lett.}}

\def\04a{{2004 a}}
\def\04b{{2004 b}}

\begin{document}

\title{Evading Cauchy horizon excision in scalarized regular black holes}

\author{Ernesto Contreras}\email{ernesto.contreras@ua.es }
\affiliation{Departamento de F\'{\i}sica, Universidad de Alicante, Campus de San Vicente del Raspeig, E-03690 Alicante, Spain}
\author{Pedro Bargue\~no}\email{pedro.bargueno@ua.es}
\affiliation{Departamento de F\'{\i}sica, Universidad de Alicante, Campus de San Vicente del Raspeig, E-03690 Alicante, Spain}
\author{Arthur G. Suvorov}\email{a.suvorov@uni-tuebingen.de}
\affiliation{Theoretical Astrophysics, Institute for Astronomy and Astrophysics, University of T\"{u}bingen, 72076 T\"{u}bingen, Germany}

\begin{abstract}
Spontaneous scalarization provides a dynamical mechanism to evade the
no-hair paradigm, but it has been argued to generically eliminate the
Cauchy horizon in charged black holes. We show that this obstruction is
not universal, but instead follows from the sign-definite structure of
the Einstein-Maxwell source term. Within the $P$-dual formulation of nonlinear electrodynamics, we derive a
general condition under which the effective scalar source changes sign
between the horizons, allowing the integral constraint to be satisfied
without destroying the Cauchy horizon. This establishes that the fate of
the Cauchy horizon depends on the electromagnetic coupling and identifies candidate theories
where scalarized horizons may persist. The resulting framework
opens the possibility of studying scalarization in the interhorizon
region and its interplay with mass inflation.
\end{abstract}

\maketitle
\section{Introduction}
The no-hair paradigm is one of the cornerstones of black-hole physics. In its standard formulation, stationary, asymptotically flat black holes in general relativity are fully characterized by a small set of global charges, namely mass, angular momentum, and gauge charges \cite{Heusler:1996jaf}. Supported by a variety of uniqueness theorems, this result implies that additional degrees of freedom are radiated away during gravitational collapse, leaving no independent ``hair'' outside the event horizon. From a physical viewpoint, this rigidity raises a natural question: to what extent can realistic matter sectors or modified interactions evade these constraints? 

While a large number of hairy solutions sourced by matter fields have been found (e.g., \cite{1989JETPL..50..346V,1991PhLB..268..371D,2023CQGra..40t5021B}), merely introducing additional fields is not entirely satisfying: a physically-meaningful notion of hair requires a well-defined mechanism that dynamically supports a nontrivial configuration. 
In this sense, the emphasis shifts from the existence of solutions to the identification of robust dynamical mechanisms. 
One such mechanism is spontaneous scalarization, originally identified for neutron stars in scalar-tensor theories \cite{1993PhRvL..70.2220D} and later extended to black-hole spacetimes. In this framework, a scalar-free (``bald'') solution becomes unstable due to a negative {square of the} effective mass induced by non-minimal couplings, triggering the growth of scalar perturbations and leading to a new branch of solutions with nontrivial scalar profiles. This phenomenon has been extensively studied in models where the scalar field couples to curvature or matter invariants, including the Gauss--Bonnet \cite{Doneva:2017bvd,Silva:2017uqg} and Maxwell \cite{Herdeiro:2018wub} invariants; see Ref.~\cite{Doneva:2022ewd} for a review. More recently, spontaneous scalarization has been extended to black holes supported by nonlinear electrodynamics (NLED) \cite{Balart:2014cga,Bokulic:2025brf} and a general framework for scalarization in regular spacetimes was developed in Ref.~\cite{Contreras:2025qbx}. 

{NLED has motivational roots outside of gravity. 
Notably, the self-energy of charged particles diverges in Maxwell theory but nonlinearities can regularize this pathology \cite{bi34}.
Inclusion of QED loop-corrections (Euler-Heisenberg theory), ultraviolet cutoffs (Born-Infeld theory), or the dynamics of gauge fields in string theories (non-Maxwellian Lagrangians are required to reproduce the scattering amplitudes of open strings at low energies \cite{ft85}) naturally yield NLED models.
However, while NLED can resolve the self-energy problem \emph{and} potentially regularize black holes with arbitrarily small charge \cite{pelltor69}, it was recently shown \cite{dt25} that NLED black holes in general relativity are generically unstable to non-radial perturbations. 
As scalar fields arguably represent the least offensive extension to general relativity one may consider, it is worth asking whether their inclusion can rescue the ability for NLED to regularize singularities.
}

{In this context,} an important tension has emerged: there is increasing evidence that spontaneous scalarization may destabilize or even eliminate the Cauchy horizon in charged or regular black holes \cite{An:2021std,Hale:2025ech}. In particular, the existence of an Cauchy horizon is closely tied to the regularity of the spacetime \cite{2018JHEP...07..023C,2024PhRvD.109j4032O,2025arXiv250703581B}. {In this work, the word \emph{regular} refers to a spacetime in which all curvature invariants remain finite and the metric extends smoothly to a regular center. Within this class of geometries, the lapse function must take positive values as $r\to0$, implying the existence of a Cauchy horizon because the lapse is also positive outside the event horizon but goes to zero there: to return to a positive value as $r \to 0$ it must therefore cut the axis again. Regularity plays a central role in our analysis since the integrated scalar-field constraint is formulated under the assumption that both the event and Cauchy horizons belong to a regular spacetime.}
As such, destabilizing the Cauchy horizon indicates that members of the scalar branch are no longer regular.
Understanding such an interplay is therefore central to any attempt to construct scalarized, \emph{regular} black holes.

In this work, we investigate the conditions under which spontaneous scalarization can be compatible with the existence of a Cauchy horizon in regular black holes. 
Our goal is to identify the class of regular black-hole models for which scalarization does not automatically excise the Cauchy horizon, thereby opening the possibility of studying the inter-horizon dynamics and the associated mass-inflation instability.
{In that respect, our results may be seen in a no-go context: the space of scalarized NLED theories that could plausibly allow for regular black holes is even more restricted.}

We begin in Section~\ref{sec:pdual} by introducing the $P$-dual formalism of NLED to construct black holes regularized by charge. The process of coupling a scalar field is then detailed in Section~\ref{sec:themodel}. We then revisit the obstruction with respect to Cauchy horizon excision that arises in the Einstein--Maxwell case, emphasizing that it follows from the sign-definite structure of the effective scalar source.
We show that this obstruction can be relaxed in NLED because the corresponding source need not preserve a definite sign (Section~\ref{sec:excision}).
Focusing on electrically-charged configurations, we derive a necessary condition for the survival of the Cauchy horizon and illustrate the mechanism with an explicit model for which the effective source admits an analytic zero.

\section{$P$-dual formalism} \label{sec:pdual}
We consider static, spherically symmetric spacetimes described by the line element
\begin{equation}\label{eq:metric}
    ds^2=-N(r)e^{-2\delta(r)}dt^2+\frac{dr^2}{N(r)}+r^2 d\Omega^2,
\end{equation}
where $N(r)$ and $\delta(r)$ are radial functions. Instead of fixing an NLED model and solving for the geometry, one may adopt an inverse approach in which the metric functions are specified \emph{a priori} and the corresponding electromagnetic theory is reconstructed. This strategy is naturally implemented within the $P$-dual (Hamiltonian) formalism \cite{Pellicer:1969,Plebanski:1987}, and has proven particularly useful for constructing regular black-hole solutions \cite{Ayon-Beato:1998hmi,Ayon-Beato:1999kuh,Dymnikova:2015hka}. Physical viability requires, in particular, the existence of a Maxwellian weak-field limit and the satisfaction of suitable energy conditions \cite{Curiel:2014}.

We consider Einstein gravity minimally coupled to a nonlinear electromagnetic field described by the action
\begin{equation} \label{eq:bareaction}
S=\int d^4x \sqrt{-g} \left[R-\mathcal{L}_{\rm EM}(F) \right],
\end{equation}
where $\mathcal{L}_{\rm EM}$ is a function of the invariant
$F=\tfrac{1}{4}F_{\mu\nu}F^{\mu\nu}$, with
$F_{\mu\nu}=\partial_\mu A_\nu-\partial_\nu A_\mu$.

Variation with respect to the metric yields the Einstein equations
\begin{equation} \label{eq:einstein}
R_{\mu\nu}-\frac{1}{2}g_{\mu\nu}R=8\pi T_{\mu\nu},
\end{equation}
with stress-energy tensor
\begin{equation} \label{stressenergy}
T_{\mu}^{\ \nu}
=
-2\mathcal{L}'(F)F_{\mu\alpha}F^{\nu\alpha}
+\frac{1}{2}\delta_\mu^{\ \nu}\mathcal{L}_{\rm EM}.
\end{equation}
Variation with respect to $A_\mu$ leads to the generalized Maxwell equations
\begin{equation}
\nabla_\mu\!\left[\mathcal{L}'(F)F^{\mu\nu}\right]=0,
\end{equation}
or, equivalently,
\begin{equation}\label{eofM}
\partial_\mu\!\left[\sqrt{-g}\,\mathcal{L}'(F)F^{\mu\nu}\right]=0.
\end{equation}

It is convenient to introduce the auxiliary antisymmetric tensor
\begin{equation} \label{eq:ptensor}
P_{\mu\nu}=\mathcal{L}'(F)\,F_{\mu\nu},
\end{equation}
in terms of which Eq.~\eqref{eofM} takes the form of the standard Maxwell equations in terms of $P$. This structure motivates the definition of the Hamiltonian density
\begin{equation} \label{eq:legendre}
\mathcal{H}(P)=2F\mathcal{L}'(F)-\mathcal{L}_{\rm EM},
\end{equation}
which is naturally viewed as a function of the invariant
$P=\tfrac{1}{4}P_{\mu\nu}P^{\mu\nu}$.

Via a Legendre transformation, we have
\begin{equation} \label{eq:lagrangian}
\mathcal{L}_{\rm EM}=2P\mathcal{H}_P-\mathcal{H},
\end{equation}
and
\begin{equation} \label{eq:faraday}
F_{\mu\nu}=\mathcal{H}_P P_{\mu\nu},
\end{equation}
where $\mathcal{H}_P=d\mathcal{H}/dP$. The stress-energy tensor \eqref{stressenergy} becomes
\begin{equation} \label{eq:stresstensor}
T_{\mu\nu}
=
\frac{1}{4\pi}\mathcal{H}_P P_{\mu\alpha}P_\nu^{\ \alpha}
-\frac{1}{4\pi}g_{\mu\nu}\left(2P\mathcal{H}_P-\mathcal{H}\right).
\end{equation}
Note that the Maxwell limit is recovered for $\mathcal{H}_P=1$.

A key advantage of the $P$-dual formalism is that the Einstein equations can be inverted to determine the electromagnetic theory compatible with a given geometry. Writing
\begin{equation}
N(r)=1-\frac{2m(r)}{r},
\end{equation}
one finds that an exact solution is obtained if
\begin{equation} \label{eq:invert}
\mathcal{H}(P(r))=-\frac{m'(r)}{r^2}.
\end{equation}
This relation provides a direct map between the geometry and the underlying nonlinear electrodynamics.
{The electric and magnetic charges are defined by integrating the coordinate-independent flux and its Hodge dual over any closed two-sphere enclosing the source, i.e.,
\begin{equation}
Q_e=\frac{1}{4\pi}\int_{S^2}\mathcal{L}_F{}^\star F,
\qquad
Q_m=\frac{1}{4\pi}\int_{S^2}F,
\end{equation}
which naturally generalize the Maxwell versions \cite{Plebanski:1987}. For the static, spherically symmetric configurations considered here, these expressions reduce to
\begin{equation}
Q_e=r^2\mathcal{L}_F F^{tr},
\qquad
Q_m=F_{\theta\phi}\csc\theta.
\end{equation}}

In the general dyonic case, the electromagnetic invariant satisfies
\begin{equation} \label{eq:generalfarad}
F(r)=\frac{2}{r^4}
\left(
Q_m^2-\frac{Q_e^2}{[\mathcal{L}'(F)]^2}
\right).
\end{equation}

In the following, we restrict to purely electric configurations,
$Q_m=0$, and denote $Q_e\equiv Q$. The electric field is then given by
\begin{equation}
E(r)=\frac{Q}{r^2\mathcal{L}'(F)}.
\end{equation}
In contrast with the Maxwell case, the nonlinear dependence in the denominator allows the electromagnetic energy
\begin{equation}
U_{\rm em}\propto\int_0^\infty dr\,r^2 E(r)^2
\end{equation}
to remain finite, providing a mechanism to regularize both the central singularity and the divergence of the Coulomb field.

Given a choice of metric function $N(r)$, the system
\eqref{eq:invert}--\eqref{eq:generalfarad} determines $F(r)$ and
$\mathcal{L}_{\rm EM}$, thereby defining a consistent NLED theory associated with the spacetime. We will use this framework to analyze how regular black-hole geometries respond to the presence of a non-minimally coupled scalar field.

\section{Spontaneous scalarization model} \label{sec:themodel}

Let us consider the action
\begin{equation}\label{eq:scalaraction}
    \mathcal{S}=\int d^4x \sqrt{-g} \left[R-2\nabla_{\mu}\phi\nabla^{\mu}\phi-f(\phi)\mathcal{L}_{\rm EM} \right],
\end{equation}
where $\phi$ is a real scalar field, $\mathcal{L}_{\rm EM}$ is the Lagrangian associated with a given NLED model, and $f(\phi)$ is a coupling function that allows for the existence of scalarized branches. In order for the scalar-free solutions to exist, one must require $f_{,\phi}(0) \equiv df/d\phi|_{\phi = 0} =0$. More generally, the coupling should satisfy conditions that both prevent pathological behavior (e.g., violation of the Newtonian limit) and permit the existence of scalarized configurations. In the present context, a particularly useful set of requirements is
\begin{equation} \label{eq:bekenstein}
    f_{,\phi\phi} >0 , \qquad \phi f_{,\phi} >0,
\end{equation}
which are the sign conditions usually associated with the onset of the tachyonic instability and the existence of the scalarized branch \cite{Herdeiro:2018wub}. Strictly speaking, the second relation is required in the region where the scalar field is supported, while on the fundamental branch it becomes a pointwise sign condition.

Varying the action \eqref{eq:scalaraction} with respect to the scalar field yields
\begin{equation} \label{eq:kleing}
0 = \nabla_{\mu} \nabla^{\mu} \phi - \frac{f_{,\phi} \mathcal{L}_{\rm EM}}{4}.
\end{equation}
For a small perturbation around the scalar-free solution, $\phi \to \delta \phi$, one finds
\begin{equation}\label{eq:perturbation}
    0 = \left[ \nabla^{\mu} \nabla_{\mu} -\frac{f_{,\phi\phi}(0) \mathcal{L}_{\rm EM}}{4}\right]\delta\phi + \mathcal{O}(\delta \phi^2),
\end{equation}
which can be interpreted as a Klein--Gordon equation with effective mass
\begin{equation} \label{eq:effectivemass}
    \mu_{\rm eff}^2 = \frac{f_{,\phi\phi}(0) \mathcal{L}_{\rm EM}}{4}.
\end{equation}
Hence, if $f_{,\phi\phi}(0)\mathcal{L}_{\rm EM}<0$, the effective mass becomes tachyonic and exponentially growing modes may develop, leading to the dynamical generation of scalar hair. The same coupling responsible for the linear instability also controls the sign structure of the fully nonlinear scalar equation.

For static electric fields, it is convenient to re-express the only non-zero component of the 4-potential through $A_{t} \equiv V(r)$.
The Euler--Lagrange equation associated with $V$ is\footnote{Note an inconsequential typo in Eq.~(24) in Ref.~\cite{Contreras:2025qbx}, where the additional term $d\mathcal{L}_{\rm EM}/{dF}$ should appear in the bracket.}
\begin{equation}
\frac{d}{dr}\left( e^{\delta}f(\phi) r^2 \frac{d\mathcal{L}_{\rm EM}}{dF} \frac{dV}{dr} \right) = 0,
\end{equation}
which can be integrated as
\begin{equation}
\frac{d\mathcal{L}_{\rm EM}}{dF}\, \frac{dV}{dr} = -\frac{Q}{e^{\delta}f(\phi) r^2}.
\end{equation}
Here, $Q$ is an integration constant interpreted as the electric charge. The remaining equations of motion derived from \eqref{eq:scalaraction} are
\begin{eqnarray}
m'&=&-r^{2}\mathcal{H}f(\phi)+\frac{1}{2}r^{2}N (\phi')^{2},\\
\delta'&=&-r(\phi')^{2},\label{deltaprima}\\
\frac{d}{dr}\left( e^{-\delta} r^2 N \phi' \right)
&=& e^{-\delta} r^2 f_{,\phi}\, \mathcal{L}_{\rm EM}.
\label{eq:total_derivative}
\end{eqnarray}
The last equation will play a central role in what follows, since its integrated form constrains the existence of a regular Cauchy horizon.
Note that primes denote differentiation with respect to radius, where we drop the argument for ease of presentation.

\section{Avoiding Cauchy horizon excision} \label{sec:excision}
{The following analysis is restricted to the standard class of static, spherically symmetric regular black holes, by which we mean spacetimes possessing a regular center with finite curvature invariants. For this class of solutions, regularity implies $m(r)=\mathcal{O}(r^3)$ as $r\to0$, so that the lapse function satisfies $N(r)\to1$ at the center \cite{lan23}. Since asymptotic flatness also requires $N(r)\to1$ as $r\to\infty$, while the existence of an event horizon implies $N(r_+)=0$ with $N<0$ immediately inside it\footnote{At least assuming the horizon is non-degenerate and the surface gravity non-zero, else the lapse could form a cusp at the axis and bounce back.}, continuity guarantees the existence of at least one additional zero $r_-<r_+$ corresponding to a Cauchy horizon. This is the only aspect of regularity used in the analysis below. Consequently, if scalarization removes the Cauchy horizon, the resulting spacetime no longer belongs to this class, although weaker notions of regularity, such as integrable singularities, may still be possible \cite{Ovalle:2025ids,sb25,Ovalle:2026singularity}.}

{To determine the conditions under which the final scalarized configuration may preserve a Cauchy horizon, we proceed as follows. 
Starting from Eq.~(\ref{eq:total_derivative}) and applying the product rule, we obtain
\begin{equation}\label{esoa}
-\delta' r^2N\phi'
+
\frac{d}{dr}\left(r^2N\phi'\right)
=
r^2f_{,\phi}\mathcal{L}_{\rm EM}.
\end{equation}
Assuming that the final configuration possesses a Cauchy horizon, we substitute Eq.~(\ref{deltaprima}) into the above and integrate between the Cauchy horizon, $r_-$, and the event horizon, $r_+$, to get
\begin{equation}
\int_{r_-}^{r_+}dr\,r^3N\phi'^3
=
\int_{r_-}^{r_+}dr\,r^2f_{,\phi}\mathcal{L}_{\rm EM}.
\label{final}
\end{equation}
{Although we restrict the discussion to the standard two-horizon case for simplicity, he same derivation applies to any pair of adjacent horizons $r_{i}<r_{i+1}$ provided the metric functions remain regular throughout the intervening region.} Note that $N$ is necessarily negative throughout the interval $r\in(r_-,r_+)$ as it is positive outside of the event horizon and vanishes at $r = r_{\pm}$. 
Restricting attention to the fundamental branch and taking $\phi>0$, spontaneous scalarization requires $\phi f_{,\phi}>0$, implying $f_{,\phi}>0$. This is the case, for instance, for the coupling $f(\phi)=e^{-\alpha\phi^2}$ with $\alpha<0$.
Equation~(\ref{final}) imposes an integrated constraint relating the scalar and electromagnetic sectors. However, additional information can be extracted from the local scalar-field equation. In particular, if $\mathcal{L}_{\rm EM}<0$ everywhere and $f_{,\phi}>0$, then
\begin{equation} \label{intermediate}
\frac{d}{dr}
\left(
e^{-\delta}r^2N\phi'
\right)
<0.
\end{equation}
Integrating from either horizon and using $N(r_\pm)=0$ gives
\begin{equation}
e^{-\delta}r^2N\phi'<0.
\end{equation}
Since $N<0$ in the interval $(r_-,r_+)$ and $N>0$ outside the event horizon, it follows that
\begin{equation}
\phi'>0,
\qquad
r\in(r_-,r_+),
\end{equation}
and
\begin{equation}
\phi'<0,
\qquad
r>r_+.
\end{equation}
Therefore, the scalar field must attain a local maximum at the event horizon, implying
\begin{equation}
\phi'(r_+)=0.
\end{equation}
Let us now reconsider Eq.~(\ref{esoa}) in the form
\begin{equation}
-\delta' r^2N\phi'
+
2rN\phi'
+
r^2N'\phi'
+
r^2N\phi''
=
r^2f_{,\phi}\mathcal{L}_{\rm EM}.
\end{equation}
Evaluating this expression at $r=r_+$, where $N(r_+)=0$, yields
\begin{equation}
r_+^2N'(r_+)\phi'(r_+)
=
r_+^2f_{,\phi}(\phi_+)\mathcal{L}_{\rm EM}(r_+),
\end{equation}
with $\phi_+\equiv\phi(r_+)$. Since $\phi'(r_+)=0$, consistency requires
\begin{equation}
f_{,\phi}(\phi_+)\mathcal{L}_{\rm EM}(r_+)=0.
\end{equation}
For the fundamental branch, $\phi_+\neq0$, and therefore $f_{,\phi}(\phi_+)\neq0$. Consequently,
\begin{equation} \label{eq:lem11}
\mathcal{L}_{\rm EM}(r_+)=0,
\end{equation}
which is in contradiction with what we assumed above Eq.~\eqref{intermediate}.
As such, Cauchy horizons are incompatible with theories for which $\mathcal{L}_{\rm EM}$ is strictly negative. In particular, in the Einstein--Maxwell--scalar model of Ref.~\cite{Herdeiro:2018wub},
\begin{equation}
\mathcal{L}_{\rm EM}
=
F_{\mu\nu}F^{\mu\nu}
<0,
\end{equation}
for purely electric configurations. Therefore, such models cannot support a regular, scalarized black hole with a Cauchy horizon.
Monomial, ``power-Maxwell'' models with $\mathcal{L}_{\rm EM} = (F_{\mu \nu} F^{\mu \nu})^p$ are similarly problematic as classical black-hole solutions also tend to be sign-definite (even powers of $p$ give $\mathcal{L}_{\rm EM} >0$ which also precludes vanishing; see, e.g., Ref.~\cite{rinc18}).
This obstruction is not a property of the scalar sector alone therefore, but rather a consequence of the sign-definite nature of the Einstein--Maxwell (or more-general electromagnetic) source term. Allowing the scalar field to develop nodes removes the obstruction; however, such configurations correspond to radially excited states and are expected to be unstable \cite{Doneva:2022ewd}. Hence, for physically relevant solutions, scalarization generically tends to eliminate the Cauchy horizon.}

{We now turn to the more general case in which $\mathcal{L}_{\rm EM}$ changes sign inside the black hole. In this situation, the sign of the source term in Eq.~(\ref{eq:total_derivative}) is no longer fixed, and the local equation does not determine a unique monotonic behavior for the scalar field. As a result, positive and negative contributions coexist on both sides of Eq.~(\ref{final}), and the preservation of the Cauchy horizon depends on the balance between these competing contributions.}

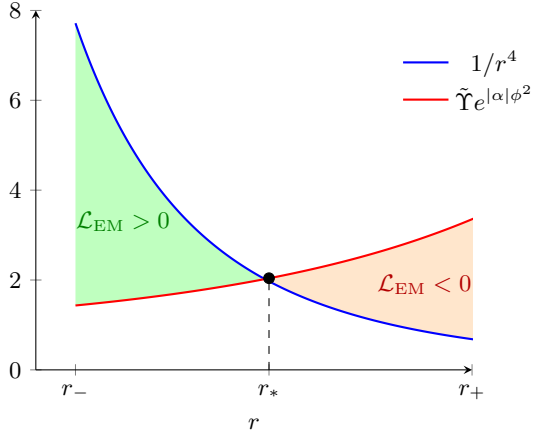
\begin{figure}[t]
\centering
\begin{tikzpicture}
\begin{axis}[
    width=0.85\linewidth,
    xlabel={$r$},
    xmin=0.55,xmax=1.1,
    ymin=0,ymax=8,
    axis lines=left,
    xtick={0.6,0.844,1.1},
xticklabels={$r_-$,$r_*$,$r_+$},
    legend style={
        draw=none,
        fill=none,
        at={(0.82,0.92)},
        anchor=north west
    },
    tick label style={font=\small},
    label style={font=\small},
]
\addplot[
    blue,
    thick,
    domain=0.6:1.2,
    samples=300
]
{1/x^4};

\addlegendentry{$1/r^4$}

\addplot[
    red,
    thick,
    domain=0.6:1.2,
    samples=300
]
{exp(x^2)};

\addlegendentry{$\tilde{\Upsilon}e^{|\alpha|\phi^2}$}

\addplot[
    draw=none,
    domain=0.6:0.84,
    samples=300,
    name path=A
]
{1/x^4};

\addplot[
    draw=none,
    domain=0.6:0.84,
    samples=300,
    name path=B
]
{exp(x^2)};

\addplot[
    green!25
]
fill between[
    of=A and B
];

\addplot[
    draw=none,
    domain=0.84:1.2,
    samples=300,
    name path=C
]
{1/x^4};

\addplot[
    draw=none,
    domain=0.84:1.2,
    samples=300,
    name path=D
]
{exp(x^2)};

\addplot[
    orange!20
]
fill between[
    of=C and D
];

\addplot[
    only marks,
    mark=*,
    black,
    mark size=2pt
]
coordinates {(0.844,2.04)};

\draw[dashed]
(axis cs:0.844,0)
--
(axis cs:0.844,2.04);

\node[green!50!black] at (axis cs:0.66,3.3)
{$\mathcal L_{\rm EM}>0$};

\node[red!70!black] at (axis cs:1.04,1.9)
{$\mathcal L_{\rm EM}<0$};

\node at (axis cs:0.86,-0.25)
{$r_*$};

\end{axis}
\end{tikzpicture}
\caption{
Schematic illustration of the sign change of the effective electromagnetic Lagrangian.
The crossing radius $r_*$ is defined by
$
r^{-4}
=
\tilde{\Upsilon}e^{|\alpha|\phi(r)^2}
$.
For $r<r_*$ one has $\mathcal L_{\rm EM}>0$, whereas for $r>r_*$ one finds $\mathcal L_{\rm EM}<0$.
}
\label{fig:crossing}
\end{figure}

To address the issue discussed above, we consider non-
linear electrodynamics in the $P$ -dual formulation. As an example, consider the model \cite{Balart:2014cga}
\begin{equation}
\mathcal H(P)=\frac{P}{\left[1+\Upsilon(-P)^{3/4}\right]^{4/3}},
\end{equation}
which is especially instructive since the sign properties of the effective source can be analyzed exactly. {Other models can also be considered. However, the corresponding analysis is considerably more involved and does not lead to equally transparent conclusions. For this reason, we do not discuss such cases here but defer an example illustrating the additional subtleties that arise in these models to the Appendix.} Now, defining
\begin{equation}
X\equiv \Upsilon(-P)^{3/4},
\end{equation}
{where $\Upsilon$ is a positive constant}, one finds
\begin{equation}
\mathcal H(P)=P(1+X)^{-4/3},
\qquad
\mathcal H_P=(1+X)^{-7/3}>0.
\end{equation}
{
The effective electromagnetic Lagrangian becomes
\begin{equation}
\mathcal{L}_{\rm EM}
=
\frac{P(1-X)}{(1+X)^{7/3}},
\end{equation}
where
\begin{equation}
P=-\frac{Q^2}{2f^2r^4}<0.
\end{equation}
Therefore, $\mathcal{L}_{\rm EM}>0$ whenever $X>1$. Since
\begin{equation}
X=\Upsilon\frac{Q^2}{2f^2r^4},
\end{equation}
this condition can be written as
\begin{equation}
\frac{1}{r^{4}}>\frac{2}{Q^{2}\Upsilon}f^{2}   . 
\end{equation}
If $f=e^{|\alpha|\phi^{2}}$, for instance, the above condition reads
\begin{equation} \label{eq:someineq}
\frac{1}{r^4}
>
\tilde{\Upsilon}\,e^{|\alpha|\phi^2},
\end{equation}
where we have renormalized the constant through $\tilde{\Upsilon}=2/\Upsilon Q^{2}$.
Inequality \eqref{eq:someineq} has a simple physical interpretation. The left-hand side
is a purely geometric radial factor, whereas the right-hand side depends
on the scalar profile through $\phi=\phi(r)$. Therefore, the two sides may
cross at some radius $r^*$, defined by
\begin{equation}
\frac{1}{(r^*)^4}
=
\tilde{\Upsilon}\,e^{|\alpha|\phi^2(r^*)}.
\end{equation}
At this point $X(r^*)=1$ and $\mathcal{L}_{\rm EM}$ necessarily changes sign. Thus,
unlike in the Einstein--Maxwell case, the nonlinear electromagnetic
sector can produce regions with $\mathcal{L}_{\rm EM}>0$ inside the black
hole.
If such a region gives the dominant contribution to the right-hand side
of Eq.~(\ref{final}), the integrated constraint can be satisfied even when
$\mathcal{L}_{\rm EM}$ is not positive throughout the whole interval
$(r_-,r_+)$ as illustrated in Figure~\ref{fig:schematic}.
\begin{figure*}
\centering
  \includegraphics[width=\textwidth]{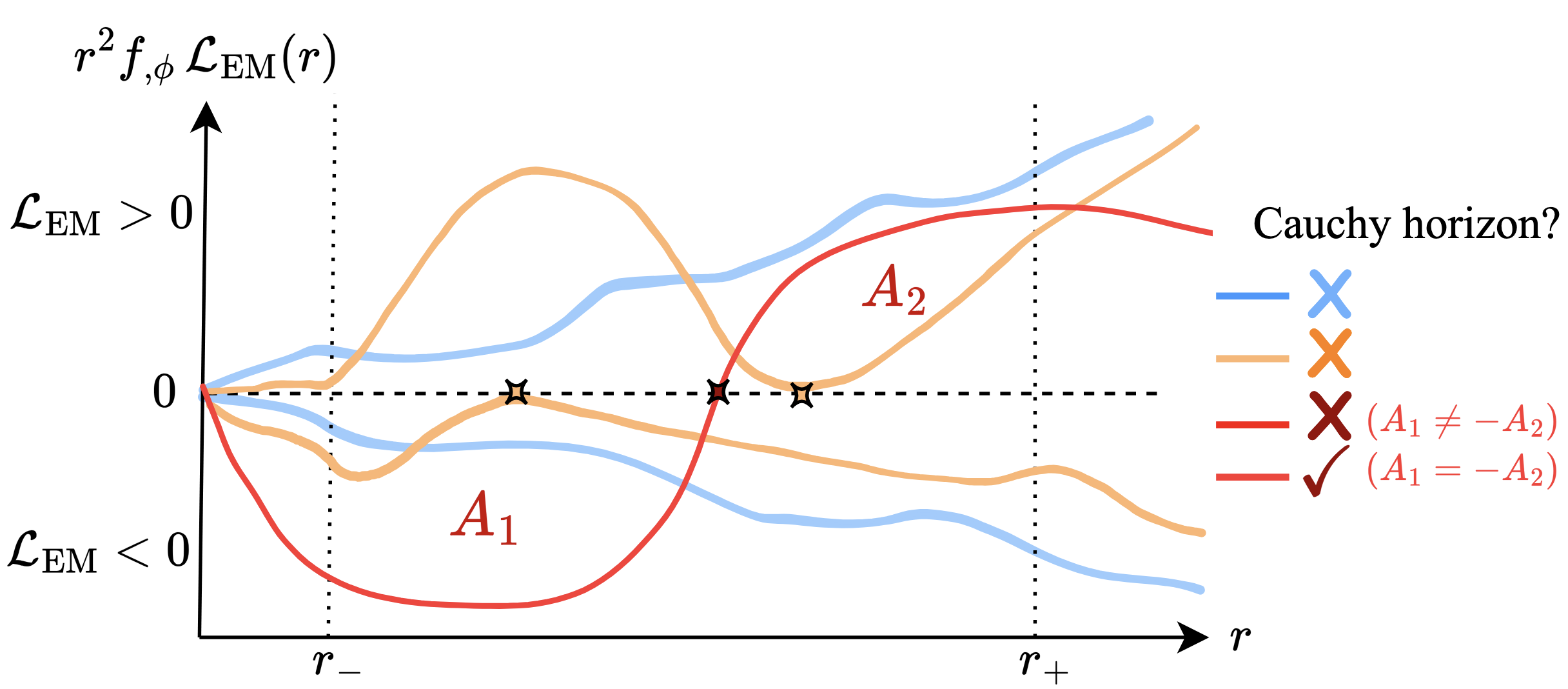}
  \caption{Sketched profiles for the integrand $r^{2} f_{,\phi}\,\mathcal{L}_{\rm EM}(r)$ and its relationship with condition \eqref{eq:lem11}, necessary for the existence of a Cauchy horizon.
  If $\mathcal{L}_{\rm EM}(r)$ is sign definite (blue), as in the Maxwell case for example, the constraint cannot be satisfied because the other factors are also sign-definite for nodeless (i.e., stable) scalar profiles thanks to the Bekenstein conditions \eqref{eq:bekenstein}.
  The existence of zeros is similarly insufficient (orange), and it is necessary not only that $\mathcal{L}_{\rm EM}$ flips sign (red) but that signed areas bounding the positive and negative portions match.
  }
  \label{fig:schematic}
\end{figure*} 
 In this sense, the crossing between $1/r^4$ and
$\tilde{\Upsilon}e^{|\alpha|\phi^2(r)}$ provides the mechanism by which
the Cauchy horizon can be preserved.
Such a condition can naturally be satisfied in the black-hole interior. Indeed, while the factor $1/r^4$ grows toward the center, the exponential contribution remains finite for regular scalar profiles. Consequently, there may exist a radius $r^{*}$ at which $X(r^{*})=1$, with $X>1$ for $r<r^{*}$ and $X<1$ for $r>r^{*}$ as shown in Fig. \ref{fig:crossing}. }

Thus, this result  provides a particularly natural candidate for preserving a scalarized Cauchy horizon.
Note that starred quantities correspond to the values at the point of sign change of $\mathcal{L}_{\rm EM}$.

At this point, some comments are in order. First, a similar strategy, based on integrating the scalar field equation across the inter-horizon region, has been employed in a series of works to establish no-inner-horizon theorems for scalarized black holes. Early results in this direction were obtained in Ref.~\cite{Cai:2020wrp}, where it was shown that static black holes with charged scalar hair do not admit a Cauchy horizon under very general conditions, independently of the scalar potential and asymptotic structure. This result was subsequently extended to theories with non-minimal derivative couplings in Ref.~\cite{Devecioglu:2021xug}, where similar conclusions were reached in Einstein--Maxwell--Horndeski models. More recently, further generalizations have been considered in Ref.~\cite{An:2021std}, where a broad class of gravitational theories with charged scalar fields was shown to exhibit the same obstruction, and in Ref.~\cite{Devecioglu:2023hmn}, where higher-curvature corrections such as Gauss--Bonnet terms were included.  

Second, the condition above is \emph{necessary}. While a sign change in 
$\mathcal{L}_{\rm EM}$ allows the integral constraint to be satisfied, it does not by itself guarantee the regularity of the resulting spacetime. In particular, preserving the Cauchy horizon requires formulating the scalarization problem in the inter-horizon region and imposing regularity at both $r_-$ and $r_+$. 

{Finally, this mechanism is intrinsically tied to the electric sector. In purely magnetic configurations, the Lagrangian description remains well-defined and invertible, and the $P$-dual formulation is not required. Accordingly, the competition between $\mathcal{H}$ and $\mathcal{H}_P$ that underlies the sign change of $\mathcal{L}_{\rm EM}$ (see Eq.~\ref{eq:legendre}) is absent, and the present mechanism does not directly extend to magnetically charged solutions. We have therefore focused on purely electric configurations, for which the second electromagnetic invariant,
\begin{equation}
G=\frac14 F_{\mu\nu}{}^\star F^{\mu\nu},
\end{equation}
vanishes identically. Consequently, although more general  theories such as Euler--Heisenberg or Born--Infeld may depend on both electromagnetic invariants, their second-invariant dependence is inactive for the class of backgrounds considered here. Restricting the analysis to theories depending only on $F$ therefore captures the relevant physics underlying the mechanism discussed in this work.}

It should be emphasized that the present analysis identifies a structural criterion for the survival of the Cauchy horizon under scalarization. Rather than being a universal outcome of the instability, the disappearance of $r_-$ is a model-dependent feature tied to the sign properties of the effective source. This is particularly relevant since the mass-inflation instability is controlled by the surface gravity at the Cauchy horizon,
\begin{equation}
\kappa_- = \frac{1}{2}\left| N'(r_-) \right|,
\end{equation}
so that the preservation of $r_-$ provides the necessary setting to analyze the interplay between scalarization and internal stability. It is also worth noting that the onset of scalarization requires  $f_{,\phi\phi}(0)\mathcal{L}_{\rm EM}<0$, which for typical  couplings such as
$f=e^{-\alpha\phi^{2}}$ implies $\mathcal{L}_{\rm EM}<0$ around the  scalar-free configuration. However, the integral constraint demands  that the effective source changes sign between the horizons. This  suggests that the nonlinear growth of the scalar field dynamically  modifies the electromagnetic sector, allowing the Lagrangian to  evolve from the configuration triggering the tachyonic instability  to a final configuration where a sign change occurs. This provides a  natural mechanism by which scalarization may remain compatible with  the existence of a Cauchy horizon.

{The implications of any candidate model should be interpreted with some care. Bronnikov's no-go theorem applies to static, spherically symmetric solutions of Einstein gravity coupled to a single  theory [i.e., $\mathcal{L}_{\rm EM} = L(F)$] with a Maxwell weak-field limit [i.e., $\lim_{F \to 0} L(F) \sim \mathcal{O}(F)$], and shows that a regular center is incompatible with a nonzero electric charge within that class \cite{Bronnikov:2000vy}. The $P$-dual formulation does not evade this statement as a theorem about fundamental $L(F)$ theories; rather, it provides an alternative Hamiltonian description in terms of $\mathcal H(P)$, at the price that the corresponding $L(F)$ formulation may become multivalued or branch-dependent \cite{Bronnikov:2000vy}. More recently, Bronnikov argued that even regular magnetic solutions with a Maxwell weak-field limit generically suffer from violations of causality, unitarity, or dynamical stability near the center \cite{Bronnikov:2000vy}, while Russo and Townsend \cite{Russo:2026causalNLED} showed that regular charged black holes are excluded within broad classes of causal NLED.
It should be emphasized, however, that these no-go results are established for  treated as an independent matter sector, without the nonminimal scalar coupling considered here. In the present model, the electromagnetic sector appears through the combination $f(\phi)\mathcal{L}_{\rm EM}$, so that the full theory does not belong to the class of pure $L(F)$ models covered by those theorems. Nevertheless, in the limit $f(\phi)\rightarrow1$, where the scalar field decouples, the ``standard'' NLED description is recovered and the usual no-go results apply.}

{As such, the existing no-go theorems should be regarded as placing restrictions on the pure  sector, rather than on the coupled scalar--electromagnetic system considered here. Whether analogous no-go theorems exist for  nonminimally coupled to scalar fields remains, to our knowledge, an open question. From this perspective, the coupled scalar--electromagnetic system should be regarded as the fundamental theory, while the pure  sector arises only as a limiting case. Within this framework, the condition derived above identifies those models for which the scalar source ceases to be sign-definite, opening the possibility that the Cauchy horizon is not necessarily destroyed.}

\section{Conclusions} \label{sec:conclusions}
We have investigated the conditions under which spontaneous scalarization can be compatible with the existence of a Cauchy horizon in regular black-hole spacetimes. In the Einstein--Maxwell case, the integrated scalar field equation leads to a no-go result: along the fundamental scalarized branch, the effective source is sign-definite, and the corresponding integral constraint cannot be satisfied if both the event and Cauchy horizons are regular. This shows that the standard destruction of the Cauchy horizon is tied not only to the scalar sector, but also to the sign structure of the underlying electromagnetic source.

Motivated by this observation, we considered scalarization in nonlinear electrodynamics formulated in the $P$-dual picture. In this framework, the integrated scalar equation depends on the $\mathcal{L}_{\rm EM}$ whose sign is controlled by the nonlinear electromagnetic sector. We showed that a necessary condition for evading the no-go argument is that $\mathcal{L}_{\rm EM}$ change sign in the inter-horizon region. In the specific model analyzed here, this condition can be made fully explicit and yields an analytic zero of the effective source.

The main result is therefore structural: the disappearance of the Cauchy horizon under scalarization is not universal, but depends on the sign properties of the electromagnetic sector. In this sense, the $P$-dual formalism provides a concrete mechanism by which the standard obstruction can be avoided without invoking nodal scalar configurations.

Although the existence of a sign change in $\mathcal{L}_{\rm EM}$ is only a necessary condition, it identifies the class of models for which an inter-horizon formulation of the scalarization problem becomes meaningful. This is the relevant setting for investigating whether the Cauchy horizon can survive scalarization and for studying the associated mass-inflation dynamics, controlled by the surface gravity $\kappa_-$. Our results therefore motivate a full nonlinear analysis of scalarized regular black holes with boundary conditions imposed at both $r_-$ and $r_+$. On a practical level, how might one go about checking whether $\mathcal{L}_{\rm EM}$ flips sign? It is worth noting in this respect that the equations of motion for the action \eqref{eq:scalaraction} require that
\begin{equation} \label{eq:eintensor}
   \frac{2 \left( G^{\theta}_{\theta} + \nabla_{\alpha} \phi \nabla^{\alpha} \phi\right)}{f (\phi)} =  \mathcal{L}_{\rm EM},
\end{equation}
where $G^{\mu}_{\nu}$ is the mixed Einstein tensor. Now from Eq. \eqref{eq:effectivemass}, $f_{,\phi\phi}(0)\mathcal{L}_{\rm EM}<0$ to ensure spontaneous scalarizations. For any model respecting $f_{,\phi\phi}>0$ from the Bekenstein conditions \eqref{eq:bekenstein}, we thus have $\mathcal{L}_{\rm EM}<0$ necessarily. Using the solution constructed with scalar profile and metric, this negativity condition can be checked simply from the geometry by equation \eqref{eq:eintensor}. In other words, ensuring a net-zero integration can be achieved using just the metric and scalar field without appeal to the specific electromagnetic fields within the problem. Going further than this requires the construction of specific solutions however, which is beyond the scope of this paper.

{Finally, we remark that our results can be seen in the context of ``no-go'' results for NLED theories.
The generic instability identified in Ref.~\cite{dt25} for non-singular, NLED black holes in GR invites the question of whether changes to the gravitational sector may evade this issue.
We have identified a necessary condition for this purpose: invoking scalar fields to escape the result quoted only has hope in NLED
theories where sign flips in the source  occur because otherwise there cannot be a Cauchy horizon and thus regularity is likely precluded.
}

\begin{acknowledgments}
EC and PB acknowledge financial support from the Generalitat Valenciana through PROMETEO PROJECT CIPROM/2022/13. EC is funded by the Beatriz Galindo contract BG23/00163 (Spain). EC and PB has been funded by the Ministerio de Ciencia, Innovación y Universidades and the Agencia Estatal de Investigación (Project PID2025-171322NB-C21, funded by MICIU/AEI/10.13039/501100011033).
AGS acknowledges funding from the European Union's Horizon MSCA-2022 research and innovation programme ``EinsteinWaves'' under grant agreement No. 101131233 and the Deutsche Forschungsgemeinschaft through individual research grant 570901071. 
\end{acknowledgments}
\appendix

\section{Analytical complexity of canonical regular black holes}
{Within the class of NLED theories depending on the single electromagnetic invariant $P$, the condition
\begin{equation}
2P\mathcal{H}_P-\mathcal{H}=0
\end{equation}
is completely general. The  model employed in the main text was chosen because it leads to a particularly transparent analytical expression. To illustrate this point, we consider below two regular black-hole solutions constructed by Ayón-Beato and García, for which the corresponding conditions become significantly more involved.
For the first model of Ref.~\cite{Ayon-Beato:1998hmi},
\begin{equation}
\mathcal H(P)=
P\frac{1-3\alpha(-P)^{1/2}}
{\left[1+\alpha(-P)^{1/2}\right]^3}
-\frac{3\beta(-P)^{5/4}}
{\left[1+\alpha(-P)^{1/2}\right]^{5/4}},
\end{equation}
it is convenient to introduce the dimensionless variable
\begin{equation}
z=\alpha\sqrt{-P}.
\end{equation}
In terms of $z$, the condition $2P\mathcal{H}_P-\mathcal{H}=0$ takes the form
\begin{equation} \label{eq:trans}
3\beta(z+6)(z+1)^{7/4}\sqrt{\frac{z}{\alpha}}
+4z(3z-8)+4=0,
\end{equation}
where $\alpha$ and $\beta$ are some constants.
The resulting expression combines polynomial and fractional-power terms in a highly nonlinear manner, preventing a simple analytical characterization of the existence and location of its zeros. Moreover, the variable $z$ is itself an implicit function of the spacetime coordinates through
\begin{equation}
-z^2/\alpha^2 = P=-\frac{Q^2}{2r^4f(\phi(r))},
\end{equation}
so that even the determination of a critical value of $z$ does not immediately translate into an explicit analytical condition on either the radial coordinate or the scalar field. Consequently, while the transcendental equation \eqref{eq:trans} could be solved numerically without much difficulty, it does not permit the same straightforward analytical discussion as the representative model considered in the main text.}

{For the second model of Ref.~\cite{Ayon-Beato:1999kuh}, where
\begin{equation}
\mathcal H(P)=
P\left[
1-\tanh^2\left(\alpha(-P)^{1/4}\right)
\right],
\end{equation}
the same zeroing condition reduces to
\begin{equation}
\sqrt{\alpha z}\tanh\left(\sqrt{\alpha z}\right)=1.
\end{equation}
Although this equation is considerably simpler than the previous one, it still does not admit a closed-form solution in terms of elementary functions. Since its left-hand side is monotonic for positive arguments, the equation possesses a unique positive root,
\begin{equation}
\sqrt{\alpha z} \approx 1.19968,
\end{equation}
or, equivalently,
\begin{equation}
z \approx \frac{1.43923}{\alpha}.
\end{equation}
Although in this simpler case the result does not permit the same straightforward analytical discussion as the representative model considered in the main text, it demonstrates that the analysis carried out in the main text can be straightforwardly-extended to other NLED theories.
}

\end{document}